**Growth control of the oxidation state in vanadium oxide thin films**

Shinbuhm Lee,[1] Tricia L. Meyer,[1] Sungkyun Park,[1,2] Takeshi Egami,[1,3,4] and Ho Nyung Lee[1,*]

[1]*Materials Science and Technology Division, Oak Ridge National Laboratory, Oak Ridge, Tennessee 37831, United States*

[2]*Department of Physics, Pusan National University, Busan 609-735, Republic of Korea*

[3]*Department of Physics and Astronomy, University of Tennessee, Knoxville, Tennessee 37996, United States*

[4]*Department of Materials Science and Engineering, University of Tennessee, Knoxville, Tennessee 37996, United States*

*E-mail: hnlee@ornl.gov

**Abstract**

Precise control of the chemical valence or oxidation state of vanadium in vanadium oxide thin films is highly desirable for not only fundamental research, but also technological applications that utilize the subtle change in the physical properties originating from the metal-insulator transition (MIT) near room temperature. However, due to the multivalent nature of vanadium and the lack of a good understanding on growth control of the oxidation state, stabilization of phase pure vanadium oxides with a single oxidation state is extremely challenging. Here, we systematically varied the growth conditions to clearly map out the growth window for preparing phase pure epitaxial vanadium oxides by pulsed laser deposition for providing a guideline to grow high quality thin films with well-defined oxidation states of $V_2^{+3}O_3$, $V^{+4}O_2$, and $V_2^{+5}O_5$. A well pronounced MIT was only observed in $VO_2$ films grown in a very narrow range of oxygen partial pressure $P(O_2)$. The films grown either in lower (< 10 mTorr) or higher $P(O_2)$ (> 25 mTorr) result in $V_2O_3$ and $V_2O_5$ phases, respectively, thereby suppressing the MIT for both cases. We have also found that the resistivity ratio before and after the MIT of $VO_2$ thin films can be further enhanced by one order of magnitude when the films are further oxidized by post-annealing at a well-controlled oxidizing ambient. This result indicates that stabilizing vanadium into a single valence state has to compromise with insufficient oxidation of an as grown thin film and, thereby, a subsequent oxidation is required for an improved MIT behavior.

**Text**

Vanadium oxides are one of few binary oxides exhibiting intriguing strong correlation effects that are critically dependent upon the oxidation state of vanadium.[1–3] The multivalent nature of vanadium leads to the stabilization of three distinct phases: (1) $V_2O_3$ ($3d^2$, $V^{+3}$) that undergoes a first-order phase transition at $T_c \approx 150$ K from a paramagnetic metal to an antiferromagnetic insulator upon cooling, accompanied by a monoclinic distortion of its rhombohedral structure (lattice constants: $a = b = 4.95$ Å, $c = 14.01$ Å).[4,5] (2) $VO_2$ ($3d^1$, $V^{+4}$) that has a metal-insulator transition (MIT) at $T_c \approx 340$ K upon cooling, in conjunction with drastic changes in optical (from infrared reflective to transparent) and structural properties [from tetragonal ($a = b = 4.55$ Å, $c = 2.86$ Å) to monoclinic ($a = 5.74$ Å, $b = 4.52$ Å, $c = 5.38$ Å, $\beta = 122.61°$)].[6–9] In particular, the MIT near room temperature makes this phase most attractive. (3) $V_2O_5$ ($3d^0$, $V^{+5}$) that is an insulator and has an orthorhombic layered structure ($a = 11.54$ Å, $b = 3.57$ Å, $c = 4.38$ Å). Its structure makes it attractive for electrode applications in, e.g., Li-ion batteries[10,11] and actuators.[12] As was found for perovskite oxides,[13–15] it is tempting to consider these vanadium oxides as oxygen sponges, by which one can obtain a strong contrast in the physical properties by reversibly transitioning between the phases. Moreover, the redox process in vanadium oxides plays an important role for enhancing the overall performance and lifetime of many energy materials and devices, such as catalysts[16] and redox-flow batteries.[17] So far, however, a simple route to redox reactions by controlling oxygen pressure has not been well explored, whereas chemical approaches to oxidize $VO_2$ into $V_2O_5$ or to reduce into $V_2O_3$ have been reported.[16] Furthermore, the robust nature of the oxidation state of vanadium in these materials has not been well understood.

For pulsed laser deposition, which is a method frequently used for fabricating vanadium oxide thin films due to its versatility in handling a wide range of oxygen partial pressure ($P(O_2)$),[18–21] $P(O_2)$ (and substrate temperature) is one of the critical growth parameters influencing significantly film quality. The $P(O_2)$ tunes the kinetic energy of growth species in the laser-induced plume.[22,23] Accordingly, the oxygen stoichiometry of vanadium oxides can be systematically controlled by $P(O_2)$. While some perovskite oxides retain their phase irrespective of huge $P(O_2)$ variation,[22,23] it is quite expectable that the phases of vanadium oxides are very sensitive to a slight change in $P(O_2)$ since the equilibrium solid-phase diagram for the V–O system is quite complex.[24]

Here, we report the growth control of the valence state in vanadium oxides by systematically controlling $P(O_2)$ during pulsed laser deposition, followed subsequently by post-annealing under highly oxidizing conditions. We found that the room temperature MIT could be observed only in VO$_2$ thin films grown in a very narrow window of $P(O_2)$. Moreover, interestingly, the MIT behavior could be further enhanced by post-annealing in high-pressure oxygen, implying that the optimal growth pressure to obtain phase pure films with high crystallinity was not enough to fully oxidize the films. This result ultimately stresses the importance of oxygen content in VO$_2$ that plays a critical role in the sharpness and resistivity ratio of the MIT.

We epitaxially deposited 50-nm-thick vanadium oxide thin films on (0001) Al$_2$O$_3$ rhombohedral ($a = b = 4.76$ Å, $c = 12.99$ Å) substrates by pulsed laser deposition. In order to check the controllability of the oxidation state, a range of oxygen pressure was explored ($P(O_2) = 10^{-5} \sim 10^{-1}$ Torr), and the optimal substrate temperature was found to be 600 °C. We ablated a sintered VO$_2$ target by a KrF excimer laser (248 nm in wavelength) at a laser fluence of 1 J/cm$^2$. To investigate the *dc* transport properties, a physical property measurement system (Quantum Design Inc.) was used with Pt contacts in four-probe geometry. X-ray diffraction (XRD) was carried out with a four-circle high-resolution x-ray diffractometer using Cu-K$\alpha_1$ radiation.

Figure 1(a) shows XRD $\theta$–$2\theta$ scans of vanadium oxide thin films grown under various $P(O_2)$. The XRD data reveal that the phase or oxidation state of the thin films is sensitively changed by a small modification of $P(O_2)$ during the film growth. In particular, the phase pure VO$_2$ epitaxial films could only be grown in a very narrow range of $P(O_2) = 10\sim25$ mTorr. When we decreased $P(O_2)$ below 5 mTorr, the $V_2O_3$ phase was preferentially grown without any inclusion of the $VO_2$ phase. On the other hand, the growth at increased $P(O_2)$ above 30 mTorr resulted in a mixture of $V_2O_5$ and $VO_2$ phases, and further increase of $P(O_2)$ to 100 mTorr eventually resulted in the pure $V_2O_5$ phase, as summarized in Fig. 1(b). Note that, due to the structural dissimilarity between the film and substrate, we could not stabilize epitaxial $V_2O_5$ films on $Al_2O_3$ substrates. The resulting films always became polycrystalline. Nevertheless, it is worth mentioning that all three oxidation states ($V^{+3}$, $V^{+4}$, and $V^{+5}$) can be individually stabilized in a very narrow window of $P(O_2) = 2{\sim}100$ mTorr, indicating that the growth pressure has to be carefully maintained for accurate control of the valence state in vanadium oxide thin films.

The vanadium oxide films showed distinctly different transport behaviors depending on $P(O_2)$, i.e., the oxidation state or phase. Figure 2 shows the temperature dependent $\rho(T)$ of vanadium oxide thin films grown under various $P(O_2)$. The $\rho(T)$ curves in Fig. 2 can be sorted into three groups: (1) All $VO_2$ films (solid lines) grown at $P(O_2) = 10{\sim}25$ mTorr showed clear MITs at $T_c \approx 340$ K; (2) $V_2O_3$ thin films (dashed lines) revealed much broader MITs than those from $VO_2$; and (3) $V_2O_5$ films that were grown at high $P(O_2) \geq 50$ mTorr represented an insulating behavior with $\rho(T)$ monotonically changing without MIT.

In order to assess the sharpness of MIT, i.e., the quality of $VO_2$ films, we compared the full width at half maximum ($\Delta T$) of the derivatives of log$\rho$ curves for heating and cooling (see the inset in Fig. 3). As summarized in Fig. 3, $\Delta T$ decreased from 18.5 to 5.5 K upon increasing $P(O_2)$, implying that the MIT in $VO_2$ films became more pronounced with a slight increase in the oxygen pressure during the growth. The best value in our case was obtained for a film grown at $P(O_2) = 20$ mTorr. Consistently, we have found that the films' crystallinity characterized by XRD rocking curves matched well with the trend in the $\Delta T$ change. This result clearly indicates that the quality of MIT behavior is directly related to both the crystallinity and the degree of oxidation in as grown $VO_2$ films. Similar MIT characteristics were also found in high quality $VO_2$ films grown on $TiO_2$ substrates.[21]

It is tempting to conclude that the use of higher pressure is preferred for growing better-oxidized $VO_2$ films, which yield a good MIT characteristic. However, as discussed before, growing thin films in higher than 25 mTorr $O_2$ results in a valence change from $V^{+4}$ to $V^{+5}$. Thus, the use of higher oxygen pressure for growing $VO_2$ is limited by the formation of $V_2O_5$. In order to check if the oxygen content in as grown $VO_2$ films can be further increased, we *in-situ* post-annealed $VO_2$ films in oxidizing condition (see Fig. 4). This approach is useful to confirm the viability of V's oxidation state control in epitaxial films. After depositing 50-nm-thick $VO_2$ thin films at various oxygen pressures, we cooled the samples to 300, 400, or 500 °C. Once the temperature is stabilized, the films were oxidized by annealing at $P(O_2) = 100$ Torr for 10 minutes. Note that annealing for 10 minutes was long enough to reach the equilibrium oxidation state.

As shown in Fig. 4, for the case of thin films grown in $P(O_2)$ = 15 mTorr, we found interestingly that oxidized $VO_2$ films revealed a drastic change in the transport properties. The best MIT behavior could be obtained from the film annealed at a moderately high temperature (300 °C). For example, $\Delta T$ was reduced from 13.6 to 5.4 K upon oxygen annealing in case of the as grown film at 15 mTorr (see open symbols in Fig. 3), accompanied by the enhancement of films' crystallinity characterized by XRD rocking curves. This enhancement can be attributed to the removal of oxygen vacancies in $VO_2$ films during post-annealing in high $P(O_2)$. However, post-annealing at 400 °C or above deteriorated the MIT characteristic due to over-oxidation, which eventually yielded the $V_2O_5$ phase as confirmed also by a XRD study.

Finally, it should be noted that the post-oxidation process widened the growth window of high quality $VO_2$ films showing the best MIT. For the as grown films at 10, 15, and 20 mTorr, $\Delta T$ and $\Delta \omega$ were reduced to ~ 5 K and ~ 0.1°, respectively, after 100 Torr $O_2$ annealing at 300 °C (open symbols in Fig. 3). Both values are comparable to or better than those of the best MIT reported in literature.[20,21] This enhancement indicates that oxygen vacancies in $VO_2$ films can be removed at temperatures as low as 300 °C during post-annealing in high $P(O_2)$. Therefore, the proper post-oxidation process stabilizes the pure $VO_2$ phase further in a wider growth window, which has been extremely challenging during the growth due to the multivalent nature of vanadium.

In conclusion, we found that the oxidation states of vanadium oxide films can be systematically designed by varying $P(O_2)$ = 2~100 mTorr during the pulsed laser deposition. Each one of the three oxidation states ($V^{+3}$, $V^{+4}$, and $V^{+5}$) can be individually stabilized in a very narrow window. The MIT in $VO_2$ films becomes much more pronounced by removal of oxygen vacancies in $VO_2$ films through a post-annealing in high $P(O_2)$. Therefore, our work provides a good understanding on growth control of the oxidation state in V–O systems, useful for the realization of advanced electronic devices, fast catalysts, and redox-flow batteries.

**Acknowledgments**

This work was supported by the U.S. Department of Energy, Office of Science, Basic Energy Sciences, Materials Sciences and Engineering Division. One of authors (S. P.) was in part supported by NRF-Korea (2012-005940).

**Figures and figure captions**

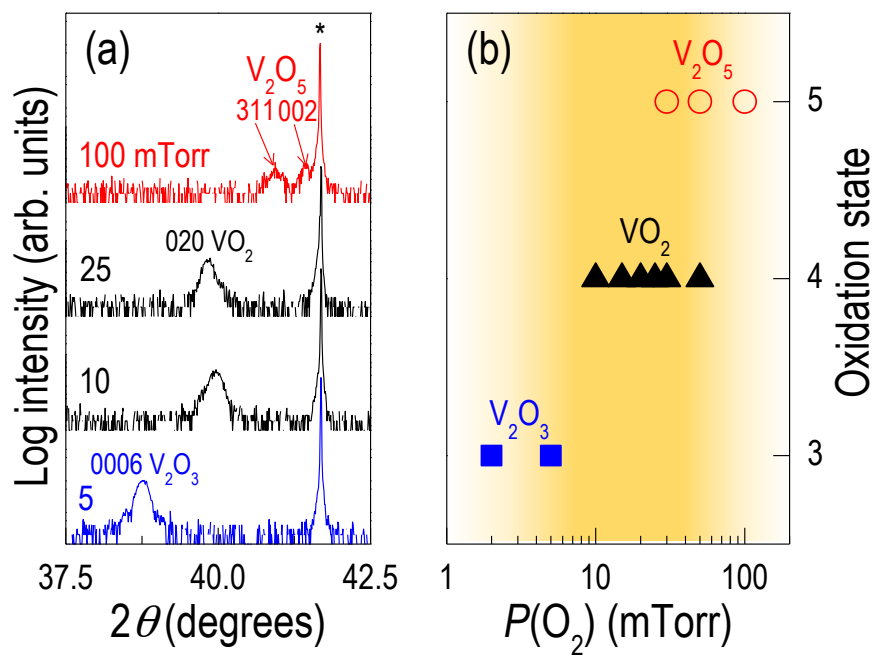

FIG. 1. (a) Selected XRD $\theta-2\theta$ scans of 50-nm-thick vanadium oxide films grown at various oxygen partial pressure $P(O_2) = 2\sim100$ mTorr. The 0006 peak of $Al_2O_3$ substrates is marked with *. (b) Preferential growth window of $P(O_2)$ for different oxidation sates ($V^{+3}\sim V^{+5}$). Solid and open symbols represent epitaxial and polycrystalline vanadium oxide phases, respectively. The polycrystalline $V_2O_5$ films were formed due to the structural dissimilarity with the $Al_2O_3$ substrate.

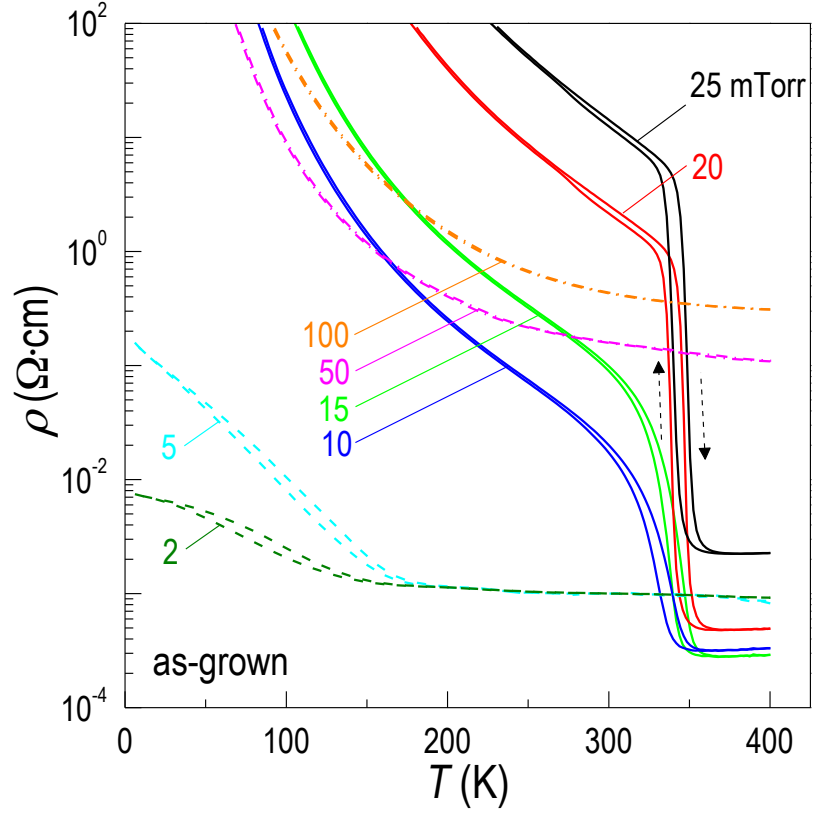

FIG. 2. $\rho(T)$ curves of as grown vanadium oxide films deposited at various $P(O_2)$. Dashed, solid, and dash-dotted lines represent three vanadium oxide phases, i.e., $V_2O_3$, $VO_2$, and $V_2O_5$, respectively.

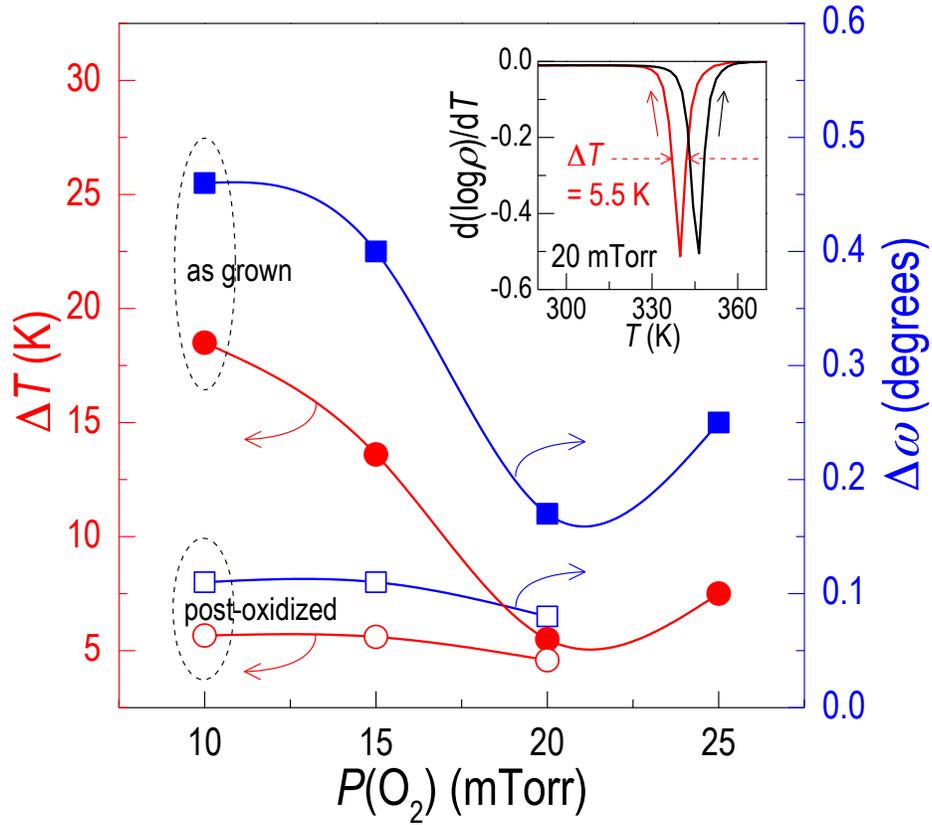

FIG. 3. Correlation between the quality of the MIT characteristic and crystallinity of VO$_2$ films. The $\Delta T$ (circles) and $\Delta \omega$ (squares) represent full width at half maximum values of the derivatives in log$\rho$ curves (see the inset) and XRD rocking curves, respectively. The solid and open symbols indicate results obtained from as grown and post-oxidized samples, respectively.

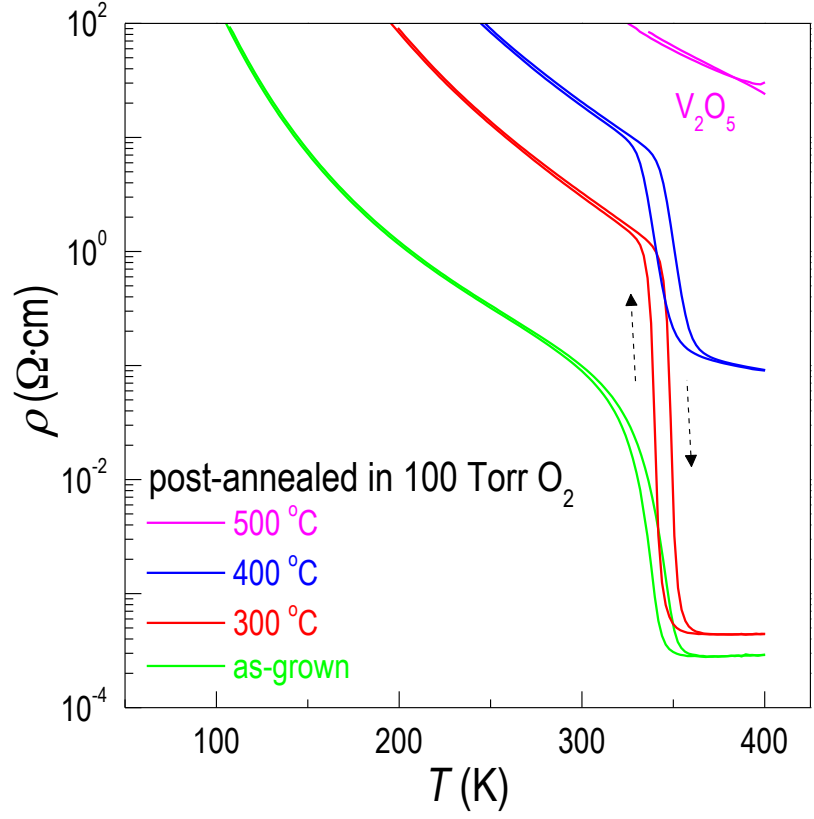

FIG. 4. $\rho(T)$ curves of VO$_2$ films *in-situ* post-annealed under a highly oxidizing condition (100 Torr O$_2$). The annealing temperature ranges from 300 to 500 °C.